\documentstyle[11pt,cs11,html,epsf]{article}

\begin{document}

\title{The CIDA-QUEST Large Scale Variability Survey in 
       the Orion OB Association: initial results}

\author{C\'esar Brice\~no and A. Katherina Vivas}
\affil{Astronomy Department, Yale University,
    New Haven, CT 06511}

\author{Nuria Calvet and Lee Hartmann}
\affil{Harvard-Smithsonian Center for Astrophysics, Cambridge, MA 02138}

\author{\& the QUEST Collaboration\altaffilmark{1}}
\altaffiltext{1}{ C. Abad, B. Adams, P. Andrews, C. Bailyn, C. Baltay,
A. Bongiovanni, C. Brice\~no, V. Bromm, G. Bruzual, P. Coppi, F. Della Prugna,
N. Ellman, W. Emmet, I. Ferr{\'\i}n, F. Fuenmayor, M. Gebhard,
R. Heinz, J. Hern\'andez, D. Herrera, K. Honeycutt, 
G. Magris, J. Mateu, S. Muffson, J. Musser, O. Naranjo,
H. Neal, G. Oemler, R. Pacheco, G. Paredes, M.  Rangel,
A. Rengstorf L. Romero, P. Rosenzweig, Ge. S\'anchez, Gu. S\'anchez,
C. Sabbey, B. Schaefer, H. Schenner, J. Shin, J. Sinnott,
J. Snyder, S. Sofia, J. Stock, J. Su\'arez D. Teller{\'\i}a,
W. van Altena, B. Vicente, K. Vieira , A. K. Vivas}

\setcounter{footnote}{3}

\begin{abstract}

Using the $8k \times 8k$ CCD Mosaic Camera on the 1m Schmidt telescope
in Venezuela, we are conducting a
large-scale, deep optical, multiepoch,  photometric (BVRIH$\alpha$) 
survey over $\sim 120\Box^\circ$ in the Orion OB association,
aimed at identifying the low mass stellar populations with ages 
$\la$ 10 Myr. 

We present initial results for a $34 \Box^\circ$ area spanning
Orion 1b, 1a and the B Cloud.
Using variability as our main selection criterion
we derive much cleaner samples than with the usual single-epoch
photometric selection, allowing us to attain a much higher efficiency 
in follow up spectroscopy and resulting in an preliminary list of
74 new low-mass ($\sim 0.4\> M_{\odot}$) pre-main sequence stars.

\end{abstract}



\section{Introduction}

Crucial aspects of theories of star formation can only be tested
by studying the stellar populations both in and near molecular clouds.
While the earliest stages of stellar evolution
must be probed with infrared and radio
techniques, many important questions can only
be studied with optical surveys of older populations with
ages $\sim$ 3-20 Myr.

In the past, studies of OB associations
have been used to investigate sequential star formation and triggering on
large scales (e.g., Blaauw 1991 and references therein).  However, because
OB stars are formed essentially on the main sequence
(e.g., Palla \& Stahler 1992) and evolve off it
in $\sim 10$ Myr, they are not useful for
infering star-forming histories on timescales of 1-3 Myr.
Moreover, it is not possible to study cluster structure and dispersal or
disk evolution without studying low-mass stars.  Studies of
individual clusters in the optical and IR have been
made (cf. Lada 1992), but these are biased toward the highest-density
regions, and cannot address older and/or widely dispersed populations.

Recent technological advances have now made it possible to carry out
large-scale studies, building on the availability of cameras 
with multiple CCDs on telescopes with wide fields of view.


\begin{figure}
\plotfiddle{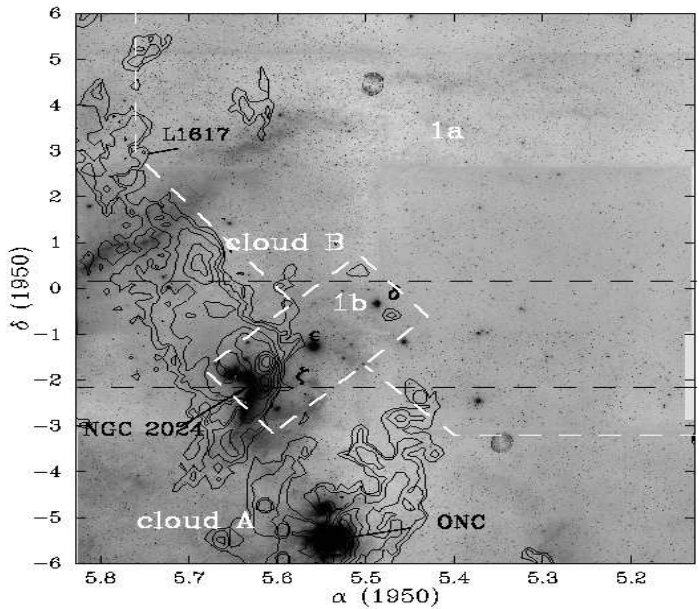}{2.5in}{0}{50.0}{40.0}{-150.0}{-70.0}
\vskip 0.2in
\caption{DSS image of the entire $120\Box^\circ$ survey area spanning the
Orion clouds and OB associations, with the molecular
complex indicated by the $^{12}{\rm CO}$ contours from Maddalena et al. (1987).
The white dashed lines outline the
associations, as described in Warren \& Hesser (1977).
The black dashed lines indicate the strip covered in our first observations,
for which we report initial results here.}
\label{fig-1}
\end{figure}

Figure 1 is shows the Orion A and B clouds and surroundings 
which we propose to survey.
The Orion belt stars, $\delta$, $\epsilon$, and
$\zeta$, are shown for reference.
The prominent bright emission nebulae are the Orion Nebula (ONC), 
NGC 2023, and NGC 2024 clusters, marking the sites of very 
recent star formation.
Also indicated are the OB associations in the region
(Blaauw 1964): Ori 1b and Ori 1a.
Ori 1d corresponds to the Trapezium/ONC region; Ori 1c is the
region surrounding it.
Photometric analysis of the O, B and A stars
(Warren \& Hesser 1977, 1978; Brown et al 1994, BGZ)
indicate ages of $< 1$ Myr (1d, see also Hillenbrand 1997), 3 Myr (1c), 
7 Myr (1b), and 12 Myr (1a).
The latest results from Hipparcos
(de Zeeuw et al 1999) yield
mean parallaxes corresponding to 330 pc (1a), 440 pc (1b), 
and 460 pc (1c), with an uncertainty of 30\%.

\begin{figure}
\plotfiddle{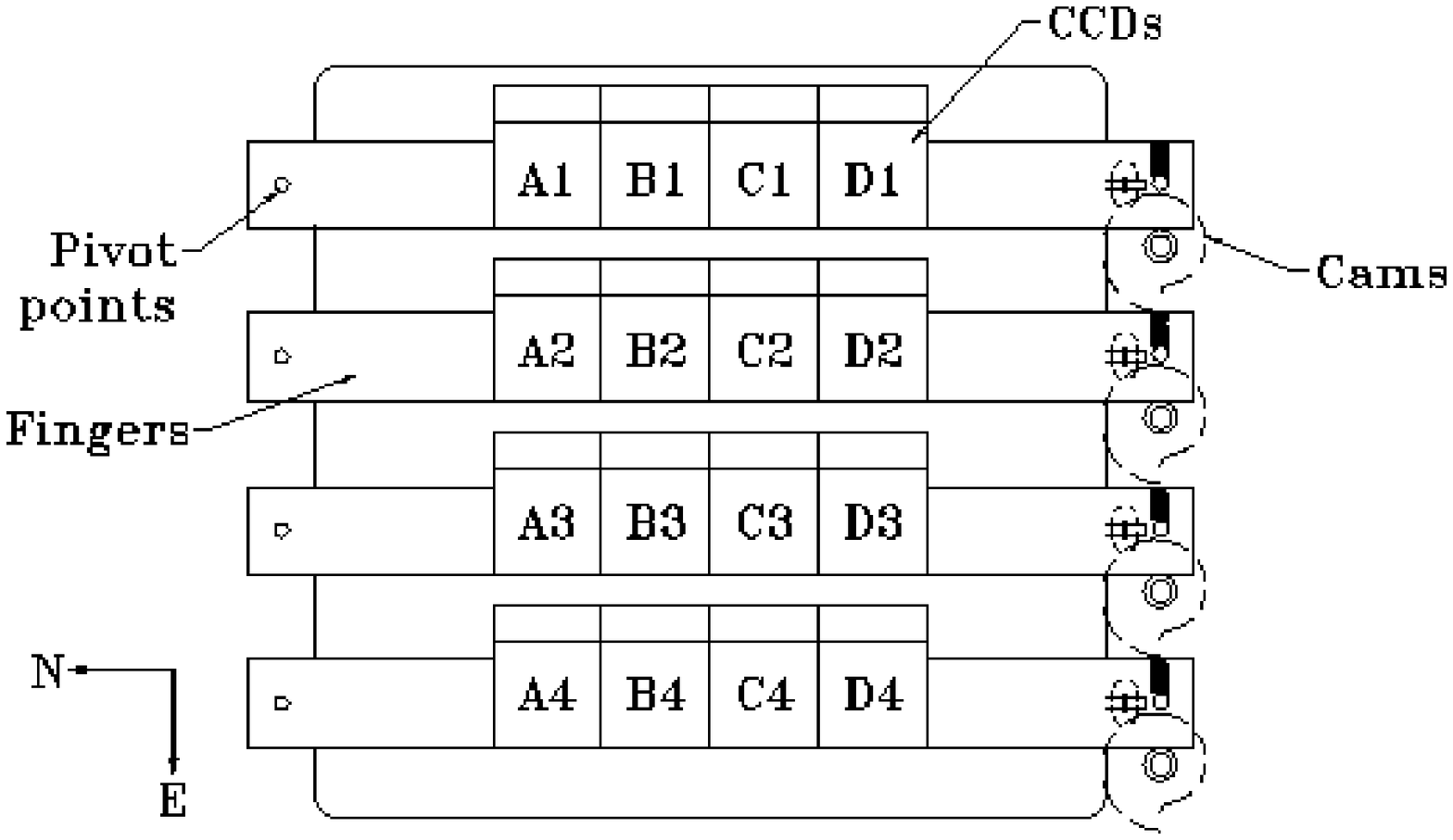}{2.5in}{0}{40.0}{40.0}{-220.0}{-20.0}
\plotfiddle{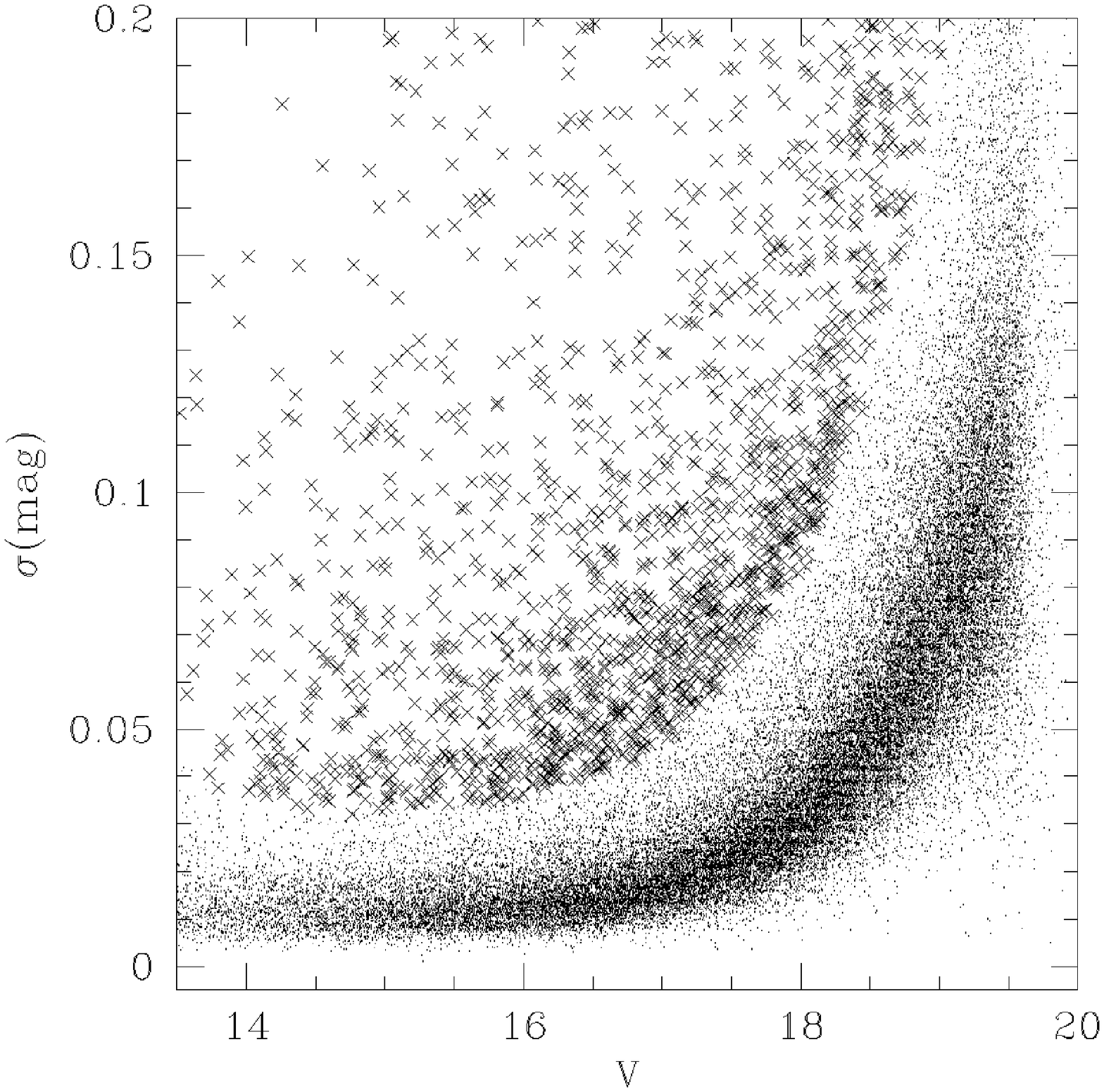}{2.5in}{0}{35.0}{32.0}{0.0}{185.0}
\vskip -3.4in
\caption{Left:  Diagram of the QuEST Camera finger assembly. Each
``finger'' of 4 CCDs has its own filter. Stars enter the array through
finger 4 and exit on finger 1. Right: The 1 $\sigma$ dispersion in the V
magnitudes in a QuEST dataset for 16 scans in Orion.
Crosses indicate variable stars (99.99\% confidence).
Most stars ($\ga 96$\%) are non-variable (dots).
}
\label{fig-2}
\end{figure}

\section{The Photometric Variability Survey}

The large scale, multiband (BVRIH$\alpha$), multi-epoch, 
deep photometric survey over $120\Box^\circ$ in Orion
(Figure 1) is being carried out
using the {QuEST\footnotemark[1]}\footnotetext[1]{ The QuEST (Quasar
Equatorial Survey Team) collaboration includes Yale
University, Indiana University,
Centro de Investigaciones de Astronom{\'\i}a, and Universidad de Los
Andes (Venezuela). The main goal of QuEST is to perform a large scale
survey of quasars.}
camera, a $8k \times 8k$ CCD mosaic detector installed on the 1m
(clear aperture)  Schmidt telescope at Llano del Hato Observatory,
in the Venezuelan Andes ($8^\circ 47'$ N, 3610 m elevation).
The 16 $2048\times 2048$ UV-enhanced, front illuminated,
Loral CCD chips are set in a 4x4 array (Figure 2, left),
covering most of the focal plane of the Schmidt telescope,
yielding a scale of 1.02'' per pixel and a field of view 
of $2.3^\circ \times 2.3^\circ$.
The camera is optimized for drift-scan observing
in the range $-6^\circ \le \delta \le +6^\circ$:
the telescope is fixed and the CCDs are read out E-W at the
sidereal rate as stars drift across the device, crossing each
of the four filters in succession.
This procedure generates a continuous
strip (or ``scan'') of the sky, $2.3^\circ$ wide; conversely, one
can survey the sky at a rate of $34.5\Box^o/hr/filter$,
down to $V_{lim}= 19.7$ ($S/N=10$).  

\subsection{Data reduction and analysis}

QuEST has developed its own software since the huge amount of data
produced is very difficult to handle with packages such as IRAF. The
whole process is completely automated with minimum interaction from
the user.  The output catalogs contain, among others,
J2000.0 positions, instrumental and calibrated magnitudes in 4 bands,
and their corresponding errors.  Positions are good to $\pm 0.2''$
down to V$\sim 19$, within a few square degrees.

\begin{figure}[h!]
\plotfiddle{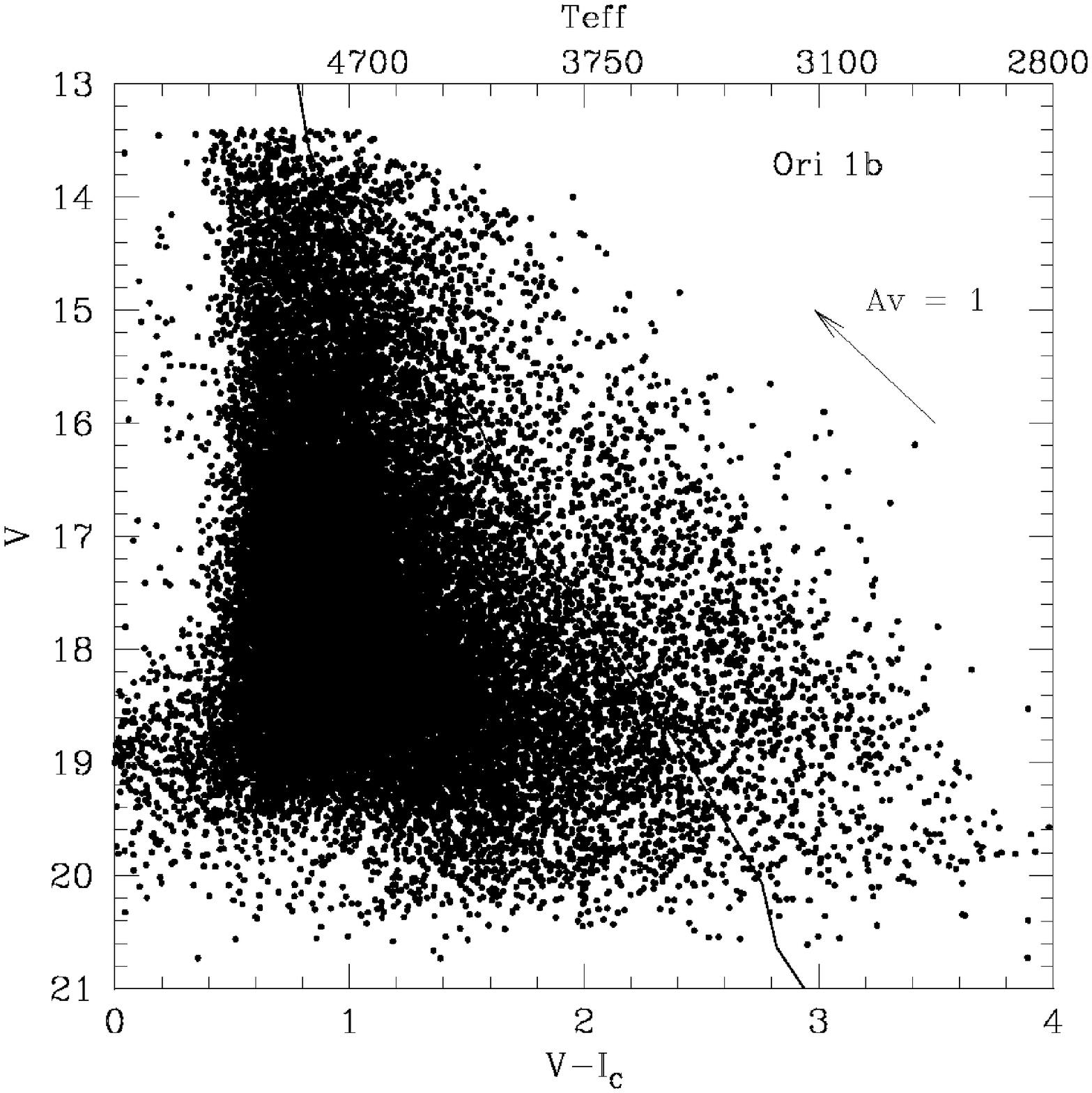}{2.0in}{0}{30.0}{30.0}{-180.0}{-35.0}
\plotfiddle{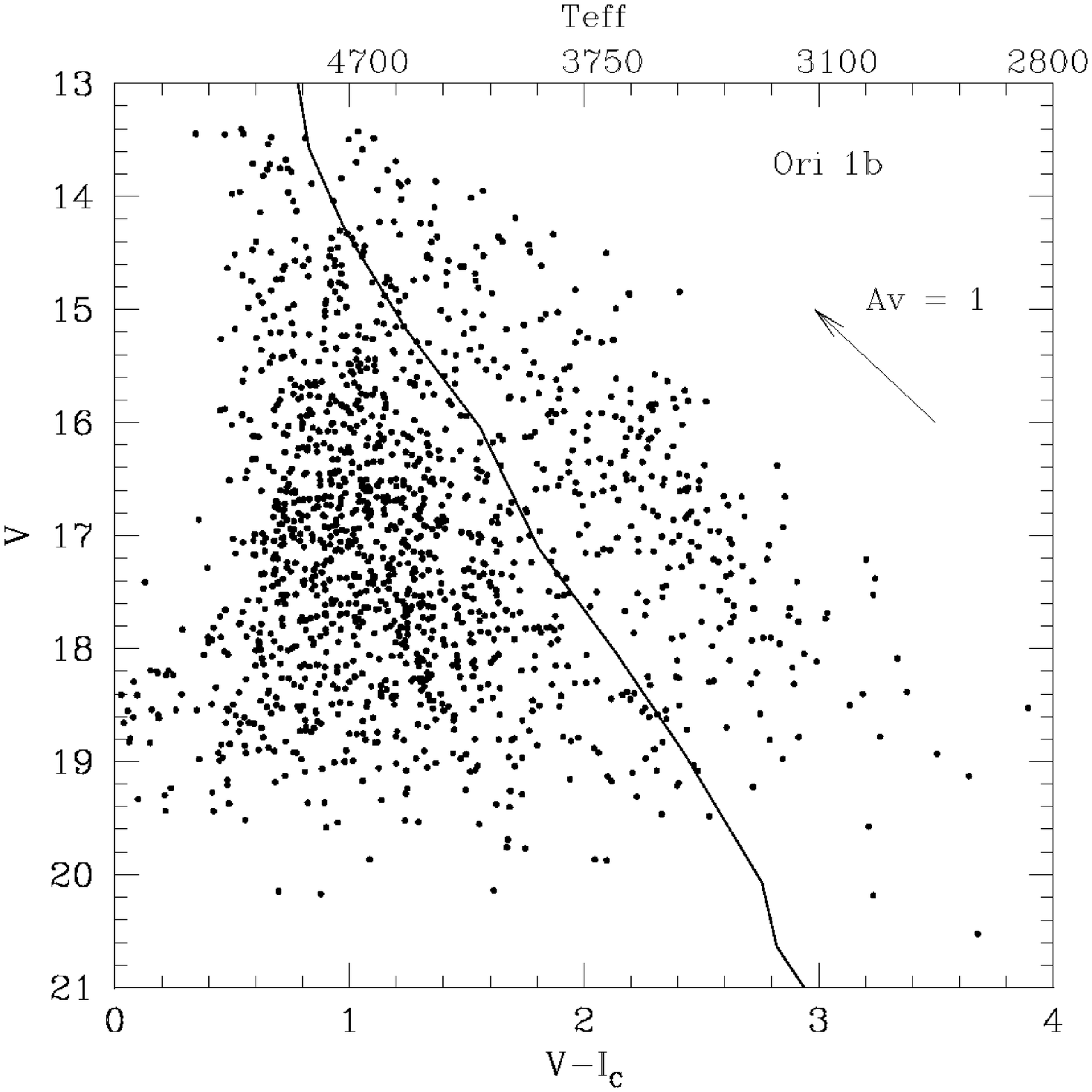}{2.0in}{0}{30.0}{30.0}{10.0}{123.0}
\plotfiddle{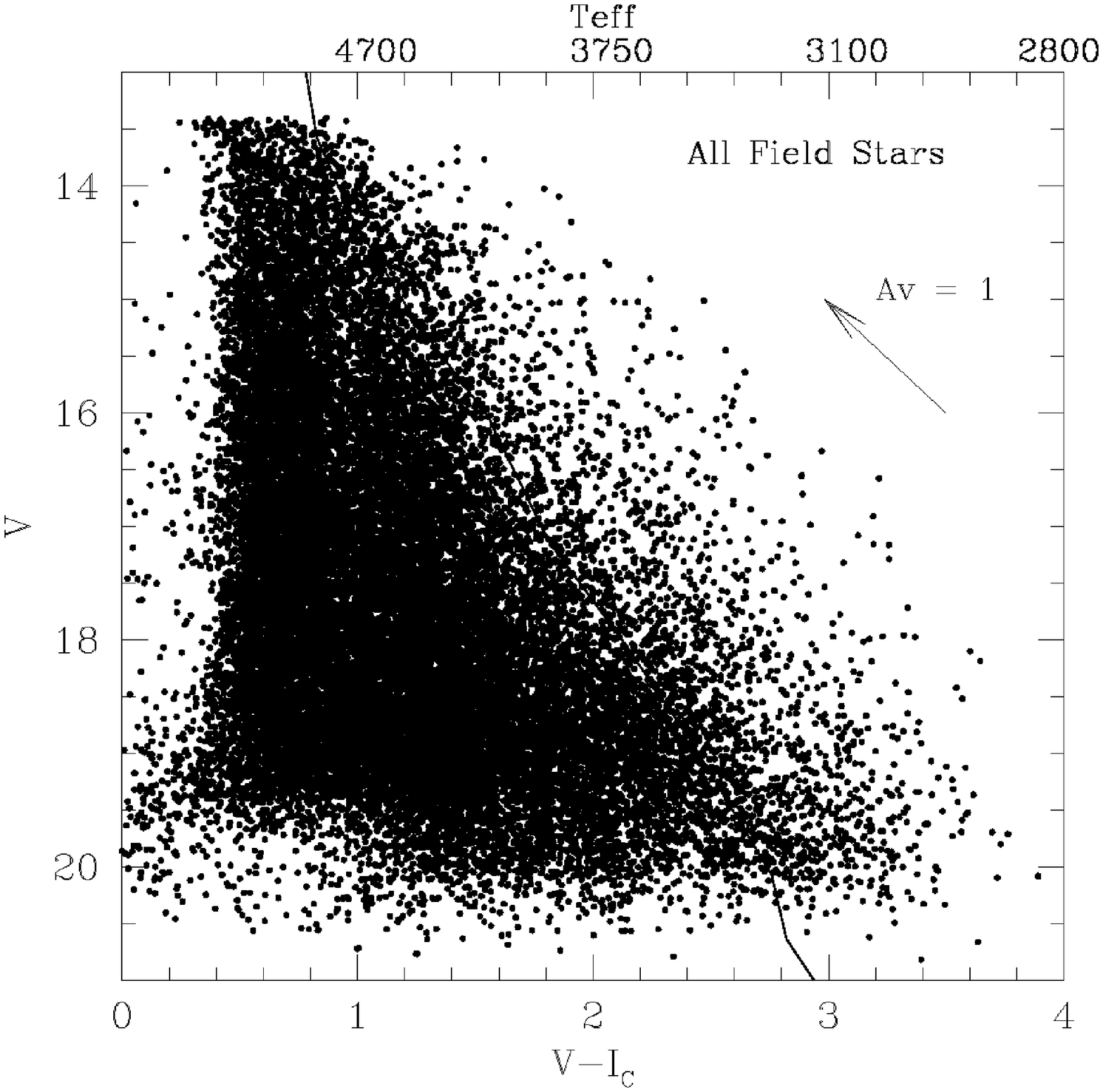}{2.0in}{0}{30.0}{30.0}{-180.0}{90.0}
\plotfiddle{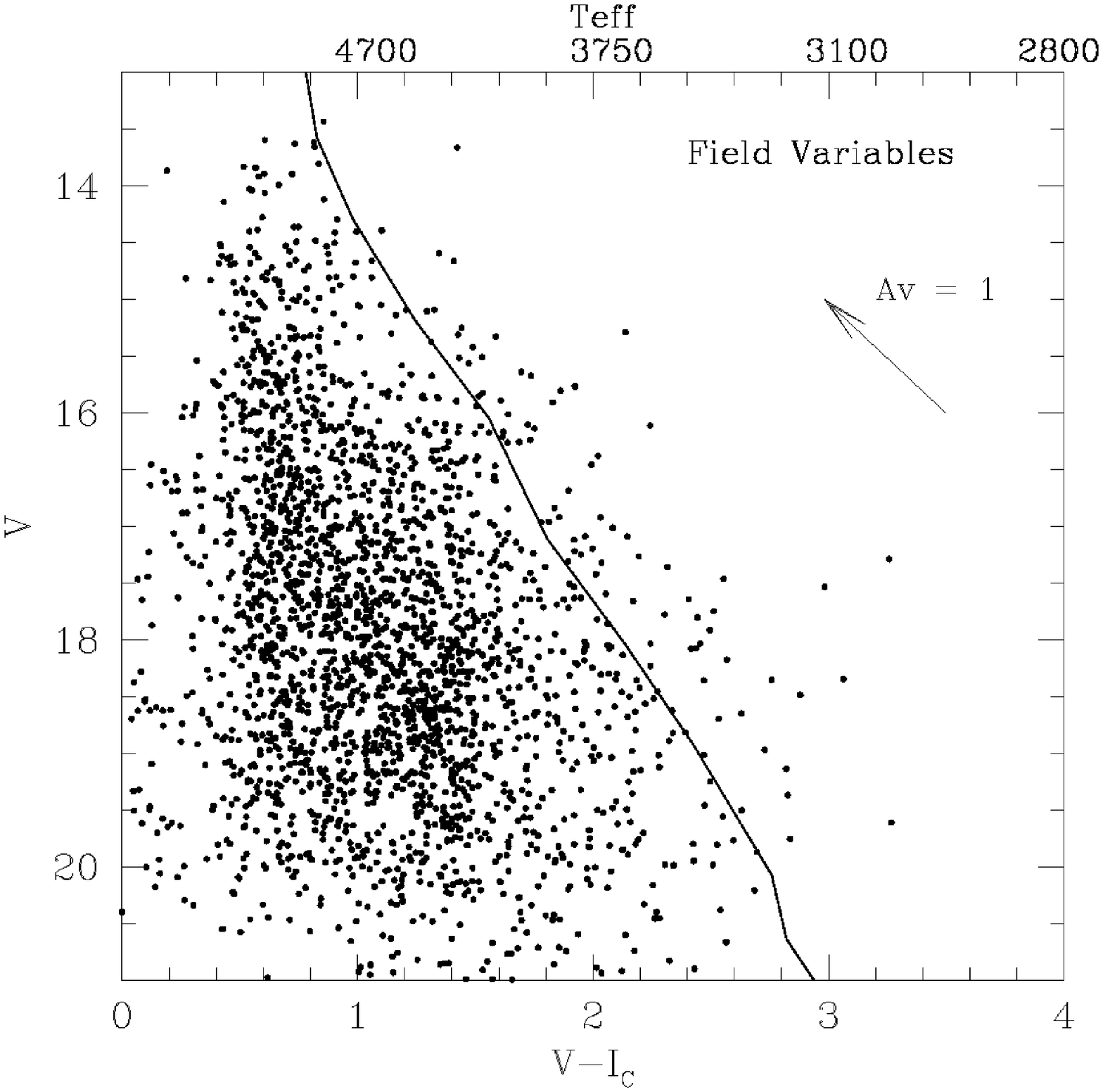}{2.0in}{0}{30.0}{30.0}{10.0}{248.0}
\vskip -4.2 true in
\caption{
V vs. V-I$_C$ diagram for different samples.
(a) Upper left: all stars in Ori 1b;
(b) upper right: all {\bf variables} in Ori 1b;
(c) lower left: all stars in a field far from the clouds;
(d) lower right: all variables in the same field.
Solid line: ZAMS.
PMS variables in 1b are clearly separated from the background,
but they do not show in the control fields.
}
\label{fig-3}
\end{figure}

\begin{figure}[h!]
\plotfiddle{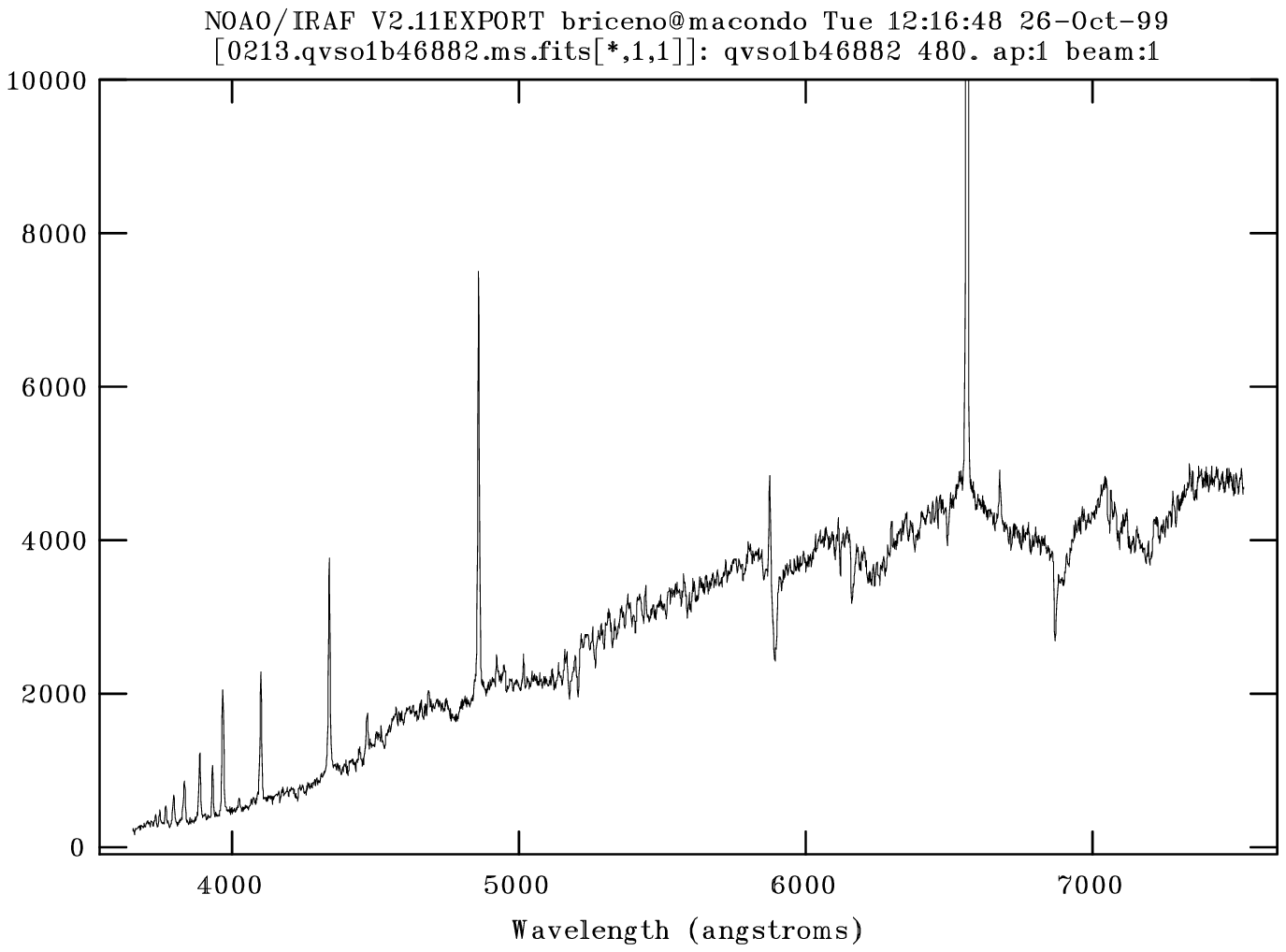}{2.0in}{0}{45.0}{40.0}{-240.0}{-55.0}
\plotfiddle{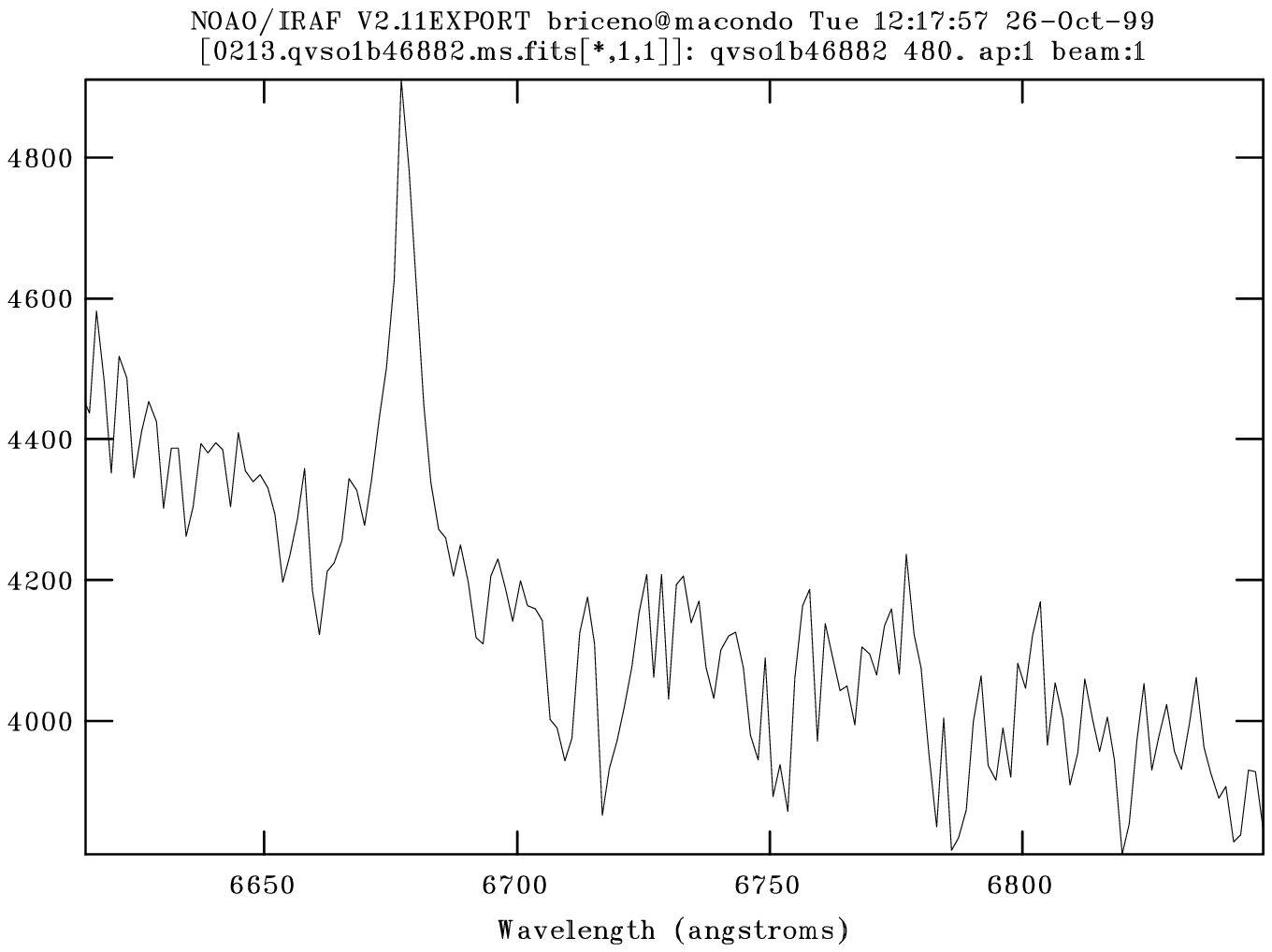}{2.0in}{0}{45.0}{40.0}{-30.0}{103.0}
\plotfiddle{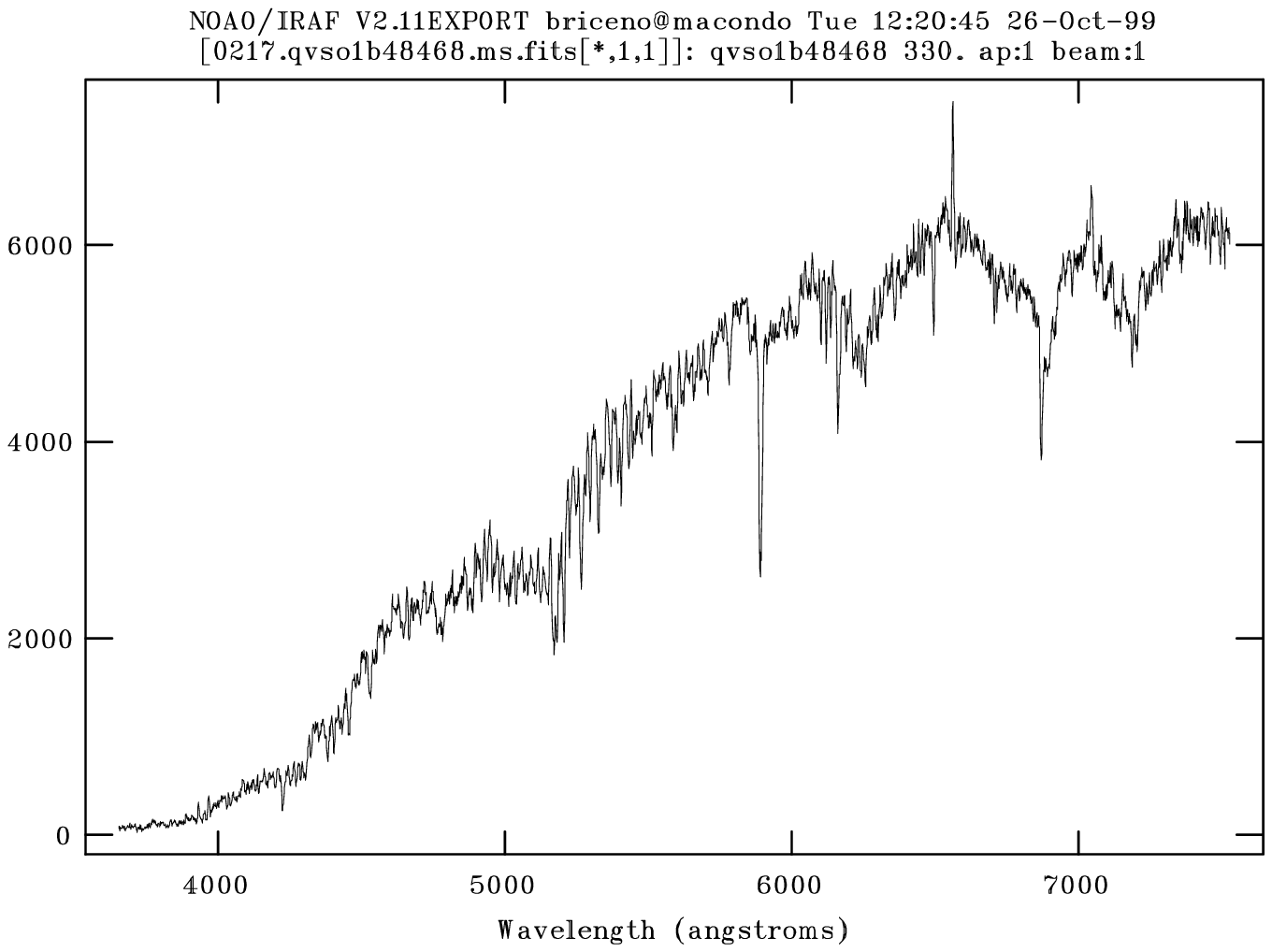}{2.0in}{0}{45.0}{40.0}{-240.0}{130.0}
\plotfiddle{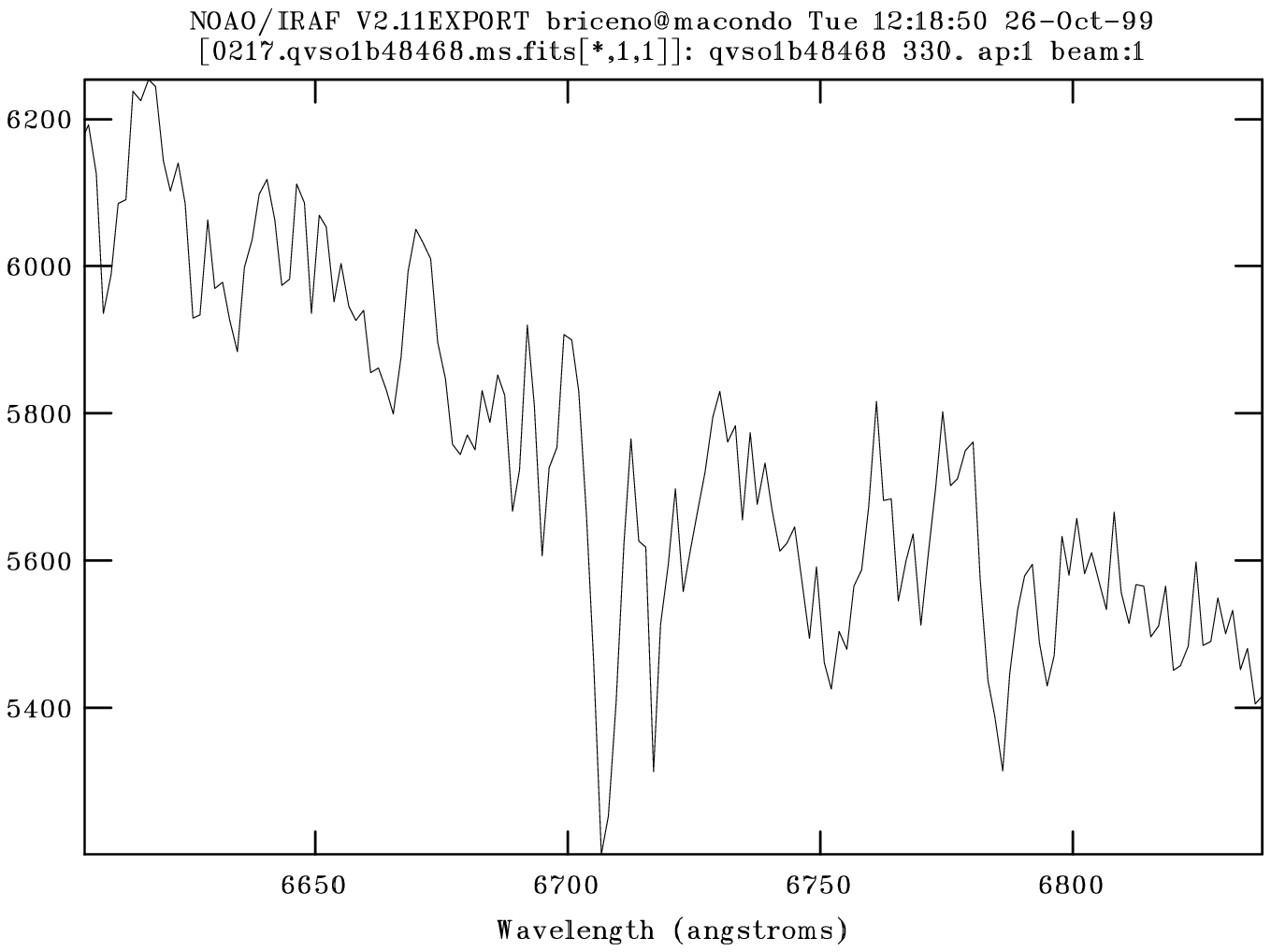}{2.0in}{0}{45.0}{40.0}{-30.0}{288.0}
\vskip -5.5 true in
\caption{
FAST spectra for 2 new M0 Orion T Tauri stars (TTS).
(a) Upper left: Classical T Tauri star (CTTS - W[H$\alpha$]= -32\AA). Other lines are
also seen clearly in emission (H$\beta$, [O I], etc). (b) Upper right: the
region around 6700\AA. The Li I$\lambda 6707$ and Ca I$\lambda 6717$
absorption lines can be clearly seen. The prominent emission is
He I$\lambda 6678$.
(c) Lower left: Weak line T Tauri star ( WTTS - W[H$\alpha$]= -1.7\AA).
(d) Lower right: region around 6700\AA, showing strong Li I$\lambda 6707$
aobsorption.
}
\label{fig-4}
\end{figure}

We have developed tools for identifying variable stars using
differential photometry.  Using a $\chi^2$ test 
and assuming a Gaussian distribution for the errors, we
consider variable only those objects for which the probability  that
the observed distribution is a result of the random errors is very
small.  In Figure 2, right, we show a sample result from the variability 
analysis in the $2.3^o$ wide strip indicated in Figure 1.
The dispersion increases for fainter objects,
so that most (non-variable) objects populate a curved region.
We use the $\chi^2$ test to identify potential variable stars
with a 99.99\% confidence level (crosses in Figure 2). 

\section{Results}

During Dec.98 - early Feb.99, we obtained 16 BVRI scans over the
strip indicated in Figure 1. We calibrated a `master' scan using
Landolt (1992) standard fields and then normalized the photometry in
all the other scans to this reference scan.

The value of variability for picking out pre-main sequence (PMS) 
candidates is shown in Figure 3. The upper left
panel is a color-magnitude diagram with all the stars
in a $\sim 10 \Box^\circ$ area within Ori 1b (the ZAMS at 440 pc 
is shown for reference);
the upper right panel shows the {\it variables} in the same
field, picked out using our selection criteria.
Populations above and below the ZAMS are
clearly separated using our variability criteria.
Indeed, the only 5 known TTS in this region (Herbig \& Bell 1988)
were all recovered as variables. We also compared
our data with the Kiso H$\alpha$ survey (c.f. Wiramihardja et al. 1993)
and found that the vast majority of H$\alpha$ stars
above the ZAMS are detected as variables ($\sim 70$\%),
but essentially none of the ones below are
(they could be a mixture of field dMe stars and
some false detections; c.f. Brice\~no et al. 1999).

To emphasize the point even further,
in the lower panels of Figure 3 we show color-magnitude
diagrams for {\it field} stars ($\alpha = 4h - 5h$, $\delta = -1^\circ$),
showing clearly that the tail of the distribution of background stars
extends far above the main-sequence, while
variables show essentially no population of PMS objects.

We have initiated followup spectroscopy of the brighter (V $\la 16.5$)
variable PMS candidates (Fig.3b),
using the FAST spectrograph (Fabricant et al. 1998)
on the 1.5m telescope at SAO, with a spectral resolution of 6{\AA } in
the range 4000 - 7000\AA.
We confirm low mass PMS stars based on
the presence of emission in H$\alpha$ and other lines, and
of Li I$\lambda 6707${\AA } strongly in absorption, which
is a reliable indicator of youth in K4-K5
and later spectral type stars (c.f. Brice\~no et al. 1997).
Even at this low resolution, Li I can be seen in late type stars
with high SNR spectra (Figure 4).
In this way, we have obtained spectra for 157 candidates and confirmed
74 of them as new TTS. We are now placing these objects in the
HR diagram to derive their masses and ages.
This high ($\sim 50$\%) efficiency is the result of the clean
selection provided by the variability criterion.

The new TTS have spectral types K7 - M2, corresponding
to masses roughly $0.5 - 0.3\>M_{\odot}$ at $\sim 1-3$ Myr.
Though preliminary, this list of new TTS already suggests
that the fraction of CTTS in 1a is much lower than in 1b, 
which would be expected if 1a is indeed older than 1b.
We are analyzing in detail the light curves of these new stars,
and spectroscopy of further candidates is under way.

This is the first optical survey to approach in spatial extent 
studies of extended star-forming regions 
(near the galactic plane, not reached by SDSS) done by other
surveys like the RASS and 2MASS, 
but going much deeper than the RASS and having
simultaneous photometry over several optical bandpasses
for many epochs, providing a unique variability database
that one-time surveys like 2MASS cannot not offer.



\begin{references}

\reference Blaauw, A. 1964, ARAA, 2, 213
\reference Blaauw, A. 1991, in The Physics of Star Formation and Early
Stellar Evolution, eds. C. Lada and N.D. Kylafis,
(Dordrecht: Kluwer), p. 125
\reference Brice\~no, C., Hartmann, L., Stauffer, J., Gagne, M., Caillault, J.-P., \& Stern, A. 1997,
\reference Brice\~no, C., Calvet, N., Kenyon, S., \& Hartmann, L. 1999, AJ, 118, 1354
\reference Brown, A.,  de Geus, E.J., \& de Zeeuw, P.T. 1994, AA, 289, 101
\reference de Zeeuw, K., Hoogerwerf, R., de Bruijne, J., Brown, A., \& Blaauw, A. 1999, AJ, 117, 354
\reference Fabricant, D., Cheimets, P., Caldwell, N. \& Geary, J. 1998, 
   PASP, 110, 79
\reference Herbig, G.H., \& Bell, K.R. 1988, Lick Obs. Bull. 1111
\reference Hillenbrand, L. 1997, AJ, 113, 1733
\reference Lada, E. 1992, ApJ, 393, 25
\reference  Maddalena, R., Morris, M., Moscowitz, J., 
   \& Thaddeus, P. 1987, ApJ, 303, 375
\reference Palla, F., \& Stahler, S.W. 1992, ApJ, 392, 667
\reference Warren, W.H., \& Hesser, J.E. 1977, ApJS, 34, 115
\reference Warren, W.H., \& Hesser, J.E. 1978, ApJS, 36, 497
\reference Wiramihardja, S., Kogure, T., Yoshida, S., Ogura, K., \& Nakano, M. 1993,
PASJ, 45, 643
\end{references}
\end{document}